\documentstyle[preprint, aps]{revtex}

\draft
\tighten

\begin{document}
\title{Quark mass density- and temperature- dependent model for bulk strange quark
matter }
\author{Yun Zhang}
\address{Department of physics, Fudan University, Shanghai 200433 , P. R. China}
\author{Ru-Keng Su$^{1,2}$}
\address{$^1$China Center of advanced Science and Technology (World Laboratory)\\
P. O. Box 8730, Beijing 100080, P. R. China \\
$^2$Department of physics, Fudan University, Shanghai 200433 , P. R. China}
\maketitle

\begin{abstract}
It is shown that the quark mass density-dependent model can not be used to
explain the process of the quark deconfinement phase transition because the
quark confinement is permanent in this model. A quark mass density- and
temperature-dependent model in which the quark confinement is impermanent
has been suggested. We argue that the vacuum energy density $B$ is a
function of temperature and satisfies $B=B_0\left[ 1-\left( 
{\displaystyle {T \over T_c}}%
\right) ^2\right] $, where $T_c$ is the critical temperature of quark
deconfinement. The dynamical and thermodynamical properties of bulk strange
quark matter for quark mass density- and temperature-dependent model are
discussed.
\end{abstract}


\section{Introduction}

It is generally accepted that the fundamental theory of strong interaction
is quantum chromodynamics (QCD), however in reality, because of its
nonperturbative characters, QCD has very little impact on the study of low
and medium energy nuclear phenomena. Many effective models, some of them
based on quark and gluon degrees of freedom \cite{1}, and the others based
on nucleons and mesons \cite{2}, or quarks and mesons \cite{3,3-1}, have
been employed to investigate the nuclear matter and strange matter. The
quark mass density-dependent model (QMDD) suggested by Fowler, Raha and
Weiner \cite{4}, is one of such candidates of effective models. Though this
model involves some arbitrary choices and can not reproduce all conclusions
given by lattice calculations, it is introduced as an alternative to the
static bag model of confinement and has substaintial application in the
study of bulk quark matter, especially strange quark matter and strange star.

Recently, since the speculation of Witten \cite{5} that the strange quark
matter (SQM) may be more stable than normal nuclei, especially, since the
argument given by Greiner and his co-workers \cite{6} that the small lumps
of strange quark matter (strangelets) may be produced in relativistic
heavy-ion collisions and could serve as an unambiguous signature for the
formation of quark-gluon plasma, much theoretical effort has been devoted to
studying the properties of SQM. Many investigations have been carried out in
the frame of MIT\ bag model \cite{7,8} or QMDD\ model. Obviously, a
successful effective model should be used to describe not only the dynamical
and thermodynamical properties of SQM, but also the phase transitions of QCD.

In this paper, we will focus our attention on the QMDD\ model. According to
the QMDD\ model, the masses of u, d quarks and strange quarks (and the
corresponding anti-quarks) are given by 
\begin{eqnarray}
m_q &=&{\frac B{3n_B}},\hspace{0.8cm}(q=u,d,\bar{u},\bar{d}),  \label{1} \\
m_{s,\bar{s}} &=&m_{s0}+{\frac B{3n_B}},  \label{2}
\end{eqnarray}
where $n_B$ is the baryon number density, $m_{s0}$ is the current mass of
the strange quark and $B$ is the vacuum energy density inside the bag. At
zero temperature,

\begin{equation}
n_B=\frac 13(n_u+n_d+n_s),  \label{su3}
\end{equation}
$n_u,n_d,n_s$ represent the density of u-quark, d-quark and s-quark,
respectively. The basic hypothesis Eqs.(\ref{1}) and (\ref{2}) in QMDD model
can easily be understood from the quark confinement mechanism. A confinement
potential which is proportional to $r$ (or $r^2$) must be added to a quark
system in the phenomenological effective models because the perturbative QCD
can not give us the confinement solution of quarks. The confinement
potential $kr$ prevents the quark goes to infinite or to the very large
regions. The large regions or the large volume means that the density is
small. This mechanism of confinement can be mimicked through the requirement
that the mass of an isolated quark becomes infinitely large so that the
vacuum is unable to support it. Thus, for a system of quarks at zero
temperature, the energy density tends to a constant value while the mass
tends to infinity, as the volume increases to infinity or the density
decreases to zero \cite{9}. This is just the picture given by Eqs.(\ref{1})-(%
\ref{su3}). In fact, the similar confinement mechanism emerges in MIT bag
model also. The boundary condition of confinement for MIT bag corresponds to
that the quark mass is zero inside the bag but infinity at the boundary or
outside the bag \cite{10}.

Although the QMDD\ model can provide a dynamical description of confinement
and explain the stability and many other dynamical properties of SQM at zero
temperature, when we extend this\ model to finite temperature and discuss
the thermodynamical behaviors of SQM, many difficulties will emerge.
Firstly, the thermodynamic potential $\Omega $ is not only a function of
temperature, volume and chemical potential, but also of density, because the
quark masses depend on density. How to treat the thermodynamics with
density-dependent particle masses self-consistently is a serious problem and
has made many wrangles for this model in references \cite{11,12,13,14}.
Secondly, as will be shown below, it can not reproduce a correct lattice QCD
phase diagram qualitatively or give us a successful equation of state when $%
n_B\rightarrow 0$. It can not describe the phase transitions of quark
deconfinement because the quark masses are independent of temperature. To
overcome this difficulty, we will suggest a quark mass density- and
temperature- dependent model (QMDTD) in this paper. Instead of a constant $B$
in Eqs.(\ref{1}) and (\ref{2}), we argue that $B$ would be a function of
temperature and choose the function $B(T)$ from Friedberg-Lee model. We will
prove that the above difficulty can be overcome in our QMDTD model.

The organization of this paper is as follows. In the next section we review
three different treatments concerning the thermodynamics with
density-dependent quark mass in refs.\cite{11},\cite{12} and \cite{13}
respectively, and prove all treatments can not overcome the second
difficulty mentioned above. In section 3, we give detailed arguments on the
temperature dependence of vacuum energy density $B$ and extend QMDD\ model
to a QMDTD model. Our results are summarized in section 4. In this section
we prove that the temperature $T$ vs density $n_B$ phase diagram for QMDTD
model becomes reasonable and it can be employed to mimic the QCD phase
transition qualitatively. The comparison of QMDD\ model and QMDTD\ model for
studying the dynamical and thermodynamical properties of SQM will also
present in section 4. The last section is a summary.

\section{Thermodynamical treatments}

At finite temperature, the antiquarks must be considered. Eq.(\ref{su3})
becomes 
\begin{equation}
n_B={\frac 13}(\Delta n_u+\Delta n_d+\Delta n_s)\hspace{0in},  \label{su4}
\end{equation}
where 
\begin{equation}
\Delta n_i=n_i-n_{\bar{i}}=\frac{g_i}{(2\pi )^3}%
\displaystyle \int %
_0^\infty d^3k\left( \frac 1{\exp [\beta (\varepsilon _i-\mu _i)]+1}-\frac 1{%
\exp [\beta (\varepsilon _i+\mu _i)]+1}\right) ,  \label{su5}
\end{equation}
($n_{\bar{i}}$) $n_i$ is the number density of the (anti)flavor $i$ ($i=u,d,s
$), $g_i=6$ is the degeneracy factor, $\mu _i$ is the chemical potential
(for antiparticle $\mu _{\bar{i}}=-\mu _i$). Inside SQM, $s$ (and also $\bar{%
s}$) quarks are produced through the weak processes 
\begin{equation}
u+d\longleftrightarrow u+s,s\rightarrow u+e^{-}+\bar{\nu}_e,d\rightarrow
u+e^{-}+\bar{\nu}_e,u+e^{-}\rightarrow d+\nu _e,  \label{su6}
\end{equation}
and similarly for antiquarks. The system of SQM must satisfy the following
constraints. The condition of chemical equilibrium yields\cite{11} 
\begin{equation}
\mu _s=\mu _d,\hspace{0.5cm}\mu _s=\mu _u+\mu _e.  \label{su7}
\end{equation}
The condition of charge neutrality reads 
\begin{equation}
2\Delta n_u=\Delta n_d\hspace{0in}+\Delta n_s+3\Delta n_e.  \label{su8}
\end{equation}
The thermodynamic potential of SQM system is 
\begin{equation}
\Omega =\sum_i\Omega _i=-\sum_i\frac{g_iT}{(2\pi )^3}%
\displaystyle \int %
_0^\infty d^3k\ln \left( 1+e^{-\beta (\varepsilon _i(k)-\mu _i)}\right) ,
\label{3}
\end{equation}
where $i$ stands for $u,d,s$ (or $\bar{u},\bar{d},\bar{s}$ ) and the
electron $e$($e^{+}$), $g_i=2$ for $e$ and $e^{+}$. Noting that $\varepsilon
_i(k)=\sqrt{m_i^2+k^2}$ and $m_{i\text{ }}$is given by Eqs.(\ref{1}),(\ref{2}%
) and (\ref{su4}), we can calculate the thermodynamic potential $\Omega $
under the constraints Eqs.(\ref{su4}),(\ref{su7}) and (\ref{su8}), and
obtain the thermodynamical quantities such as number density $n_i$,
pressure, internal energy and etc.. Due to the density-dependent quark mass,
many different treatments had been given in the references.

\subsection{First treatment}

The first thermodynamical treatment for QMDD\ model was given by Chakrabarty 
\cite{11}. After getting the thermodynamic potential $\Omega $, he used the
usual thermodynamical formula to calculate the number density $n_i$, total
pressure $p$ and the total energy density $\varepsilon $, and found 
\begin{equation}
n_i=-{\frac 1V}\left. {\frac{\partial \Omega }{\partial \mu _i}}\right|
_{T,n_B},  \label{7}
\end{equation}

\begin{equation}
p=-{\frac \Omega V},  \label{8}
\end{equation}

\begin{equation}
\varepsilon ={\frac \Omega V}+\sum_i\mu _in_i-{\frac TV}\left. {\frac{%
\partial \Omega }{\partial T}}\right| _{\mu _i,n_B}.  \label{9}
\end{equation}
After comparison with the results given by MIT bag model, Chakrabarty
claimed the properties of SQM given by QMDD model were found to be very
different from those predicted by the MIT bag model. Since the density
dependence of quark mass has not completely and explicitly taken into
account in this thermodynamical calculations, this treatment seems incorrect%
\cite{12}. But in order to compare with other treatments, we list this
treatment here also.

\subsection{Second treatment}

The second different treatment is given by Benvenuto and Lugones \cite{12}.
They claimed that the features found by Chakrabarty are consequences of an
incorrect thermodynamical treatment for QMDD model. In deriving the energy
density and the pressure, an extra term appears due to the dependence of the
quark mass on the baryon density.

The results become

\begin{equation}
p=-{\frac 1V}\left. {\frac{\partial (\Omega /n_B)}{\partial (1/n_B)}}\right|
_{T,\mu _i}=-{\frac \Omega V}+{\frac{n_B}V}\left. {\frac{\partial \Omega }{%
\partial n_B}}\right| _{T,\mu _i},  \label{10}
\end{equation}

\begin{equation}
\varepsilon ={-}p+\sum_i\mu _in_i-{\frac TV}\left. {\frac{\partial \Omega }{%
\partial T}}\right| _{\mu _i,n_B},  \label{11}
\end{equation}
and $n_i$ still satisfies Eq.(\ref{7}). The extra term produces significant
changes in the energy per baryon, make the pressure take the negative value
in the low density region and shift the stability window of strange matter
(SM). In almost all cases they found that the properties of SQM in the QMDD
model are nearly the same as those obtained in the MIT bag model.

\subsection{Third treatment}

The third different treatment is done by Peng and his coworkers\cite{13}.
Their improvements include: (1) Based upon a quark condensates argument,
instead of Eqs.(\ref{1}) and (\ref{2}), they introduce

\begin{eqnarray}
m_q &=&{\frac D{n_B^{1/3}}},\hspace{0.8cm}(q=u,d,\bar{u},\bar{d}),
\label{12} \\
m_{s,\bar{s}} &=&m_{s0}+{\frac D{n_B^{1/3}}},  \label{13}
\end{eqnarray}
where $D$ is a parameter usually determined by stability arguments. (2) They
agree with the second treatment that one must add an extra term to the
pressure formula because of the quark mass density-dependence, but do not
agree with them for adding an extra term to the expression of the energy
density because it can not give a correct QCD vacuum energy. The pressure
and the energy density given by this treatment are

\begin{equation}
p=-{\frac 1V}\left. {\frac{\partial (\Omega /n_B)}{\partial (1/n_B)}}\right|
_{T,\mu _i}=-{\frac \Omega V}+{\frac{n_B}V}\left. {\frac{\partial \Omega }{%
\partial n_B}}\right| _{T,\mu _i},  \label{14}
\end{equation}
\begin{equation}
\varepsilon ={\frac \Omega V}+\sum_i\mu _in_i-{\frac TV}\left. {\frac{%
\partial \Omega }{\partial T}}\right| _{\mu _i,n_B}.  \label{15}
\end{equation}
In fact, this treatment is a ''mixture'' of the first and the second
treatment. It chooses the pressure of the second treatment and the energy
density of the first treatment as its pressure and energy density
respectively.

Now we are in the position to study the thermodynamical behavior of SQM by
using the QMDD\ model. The temperature $T$ vs density $n_{B\text{ }}$curves
are shown in Fig.1 by three dashed lines for three treatments respectively
where we choose the parameters $B=170\mbox{MeV}\mbox{fm}^{-3},m_{s0}=150%
\mbox{MeV},D=140\mbox{MeV}\mbox{fm}^{-1}$ and $P=400\mbox{MeV}\mbox{fm}^{-3}$%
. We see from Fig.1 that the temperature T tends to infinite when $%
n_B\rightarrow 0$. This result is treatments-independent and can easily be
understood if we notice the basic hypothesis of QMDD\ model, namely, the
Eqs.(\ref{1}) and (\ref{2}) (or Eqs.(\ref{12}) and (\ref{13})), the quark
masses are divergent when $n_B\rightarrow 0$. To excite an infinite weight
particle, one must prepare to pay the price of infinite energy, i.e.
infinite temperature. This result demonstrates that the confinement in QMDD\
model is permanent. The quark can not be deconfined for any temperature.
This model can not describe the quark deconfinement phase transition and
give us a correct phase diagram of QCD.

\section{Quark mass density- and temperature- dependent model}

Obviously, if we hope to employ the QMDD\ model to mimic the phase
transition of QCD, the first problem is to avoid the permanent confinement
mechanism given by Eqs.(\ref{1}) and (\ref{2}) (or Eqs.(\ref{12}) and (\ref
{13})). It would be useful to recall what happen in MIT bag model and
Friedberg-Lee soliton bag model \cite{15}. MIT bag model is a permanent
quark confinement model because the confined boundary condition does not
change with temperature. The vacuum energy density $B$ is a constant in MIT
bag model. Contrary, the Friedberg-Lee soliton bag model is a impermanent
quark confinement model. Its confinement mechanism comes from the
interaction between quarks and a non-topological scalar soliton field. Since
the spontaneously breaking symmetry of scalar field will be restored at
finite temperature, the non-topological soliton will disappear and the quark
will deconfine at critical temperature. In this model, the vacuum energy
density $B$ equals to the different value between the local false vacuum
minimum and the absolute real vacuum minimum. This value depends on the
temperature. It means that $B$ must be a function of temperature in Eqs.(\ref
{1}) and (\ref{2}) if we hope to deconfine quark. In order to describe the
phase transition of QCD, we must extend the QMDD model to a quark mass
density- and temperature- dependent (QMDTD) model and suppose that $B$ is a
function of temperature. This is our first argument.

Our second argument comes from the calculations of the effective masses of
nucleons and mesons recently. We can sum the tadpole diagrams and the
exchange diagrams for mesons by Thermo Field Dynamics and find the masses of
nucleons and mesons all decrease with temperature \cite{16,17,18,19}. This
result for $\rho $-meson is in agreement with recent experiments \cite{20,18}%
. According to the constitute quark model, the nucleon are constructed by
three quarks and the meson by two quarks. It means that in a satisfying
quark model we must consider the temperature dependence of the quark mass.
But this effect has not been taken into account by Eqs.(\ref{1}) and (\ref{2}%
) if $B$ is a constant.

According to the conclusion of Benvenuto and Lugones \cite{12}, the results
found by QMDD\ model are nearly the same as that obtained in the MIT bag
model. On the other hand, as was pointed out by \cite{15}, the MIT bag model
can be obtained from Friedberg-Lee soliton bag model provided fixed some
parameters. Then it is natural to choose $B(T)$ given by Friedberg-Lee model
as our input.

Introducing an {\em ansatz}

\begin{eqnarray}
\ B &=&B_0\left[ 1-\left( \frac T{T_c}\right) ^2\right] ,0\leq T\leq T_c
\label{16} \\
B &=&0,\text{ }T>T_c  \label{17}
\end{eqnarray}
where $T_c$ is the critical temperature of deconfinement. When $0\leq T\leq
T_c$, Eqs.(\ref{1}) and (\ref{2}) become 
\begin{eqnarray}
m_q &=&{\frac{B_0}{3n_B}}\left[ 1-\left( \frac T{T_c}\right) ^2\right] ,%
\hspace{0.8cm}(q=u,d,\bar{u},\bar{d}),  \label{18} \\
m_{s,\bar{s}} &=&m_{s0}+{\frac{B_0}{3n_B}}\left[ 1-\left( \frac T{T_c}%
\right) ^2\right] .  \label{19}
\end{eqnarray}
The masses of quarks not only depend on the density $n_{B\text{ ,}}$ but
also on the temperature $T$. And when $T\geq T_c$, $m_q=0$, $m_{s,\bar{s}%
}=m_{s0}$. When $T=0$, Eqs.(\ref{18}) and (\ref{19}) reduce to Eqs.(\ref{1})
and (\ref{2}), and our QMDTD model reduces to QMDD model.

In our later calculations, we prefer to the thermodynamical treatment of
Eqs.(\ref{14}) and (\ref{15}), because they consider the density-dependent
masses and the QCD vacuum energy explicitely. Substituting Eqs.(\ref{18})
and (\ref{19}) into Eq.(\ref{3}), under the constraints Eqs.(\ref{su4}),(\ref
{su7}) and (\ref{su8}), and using the same argument as that of the third
treatment, we find the thermodynamical quantities $n_i$, $p$, $\varepsilon $%
, which is still expressed by Eqs.(\ref{7}),(\ref{14}) and (\ref{15})
because Eqs.(\ref{18}) and (\ref{19}) have the same density-dependence as
that of Eqs.(\ref{1}) and (\ref{2}). Even though the expressions of $n_i$, $%
p $ and $\varepsilon $ are the same, but we would like to emphasize that the
results given by our model and QMDD\ model are different because the
thermodynamical potentials calculated from Eqs.(\ref{1}) and (\ref{2}) and
Eqs.(\ref{16})-(\ref{19}) are quiet different at finite temperature. Our
results are summarized in next section.

\section{Results and discussion}

The numerical calculations have been done by adopting the parameters $B_0=170%
\mbox{MeV}\mbox{fm}^{-3}$, $m_{s0}=150\mbox{MeV}$, $T_c=170\mbox{MeV}$. The
temperature $T$ vs baryon number density $n_B$ figure is shown in Fig.1
where the pressure $P$ is fixed to be $400\mbox{MeV}\mbox{fm}^{-3}$. The
three dashed lines refer to three different treatments of QMDD models,
respectively, and the solid line refers to our QMDTD model. We see from
Fig.1 that the basic difference between QMDTD\ model and QMDD model is: when 
$n_B\rightarrow 0$, the temperature approaches to a critical temperature $%
T_c=170\mbox{MeV}$ in our model, and diverges in QMDD\ model. It means that
QMDTD model is a impermanent confinement model. It can be employed to
describe the phase transition of QCD qualitatively.

The same curves of QMDTD model but for different pressures $p=400,300,200$ $%
and$ $150\mbox{MeV}\mbox{fm}^{-3}$ are shown in Fig.2 respectively. We see
from Fig.2 that the conclusion $T$ approaches to $T_c$ when $n_B\rightarrow
0 $ will not change with pressure. The basic physical reason is that in
QMDTD model, $B$ is a monotonously decreasing function in the region $0\leq
T<T_c$ and becomes zero when $T$ approaches to $T_{c\text{ }}$. The
singularity of quark mass at zero density of QMDD model has been wiped out
in QMDTD model. When $T=T_c$, $m_{q,\bar{q}}=0$ and $m_{s,\bar{s}}=m_{s0%
\text{.}}$ It guarantees that the divergence of temperature at zero density
will not happen in QMDTD model no matter the values of pressure be.

To compare our model with QMDD model further, we investigate the
thermodynamical stability of SQM at finite temperature. The energy per
baryon vs baryon density curves at $T=50\mbox{MeV}$ for QMDTD model and QMDD
model are shown in Fig.3 where the solid line refers to QMDTD model and the
three dashed lines for three different treatments of QMDD model
respectively. We see that the solid line is lower than the others. It means
that the SQM described by QMDTD model is more stable than that by QMDD
model. The values of $n_B{}_0$ and $(\varepsilon /n_B)_0$ at the saturation
point are summarized at Table 1. We see from Table 1 that the value of $%
n_B{}_0$ for QMDTD model is situated between the maximum value $0.55\mbox{fm}%
^{-3}$ and the minimum value $0.36\mbox{fm}^{-3}$ for the second and the
third treatments of QMDD model, but the energy per baryon $(\varepsilon
/n_B)_0=1006\mbox{MeV}\mbox{fm}^{-3}$ is lower than all treatments. The
point marked with a heavy dot in the solid line is the zero pressure point
for QMDTD model, as can be seen clearly, which matches the lowest-energy
state and satisfies the basic requirement of thermodynamics pointed by ref.%
\cite{13}. The results obtained from Fig.3 represent that our model can
reproduce all thermodynamical properties of SQM which has been explained by
QMDD\ model.

The study of the equation of state for QMDTD model will show that this model
is suitable for describing the thermodynamical behavior of SQM. The curves
of pressure vs energy density are shown in Fig.4 where the solid line refers
to QMDTD model and the dashed lines for different treatments of QMDD model
respectively. We see from Fig.4 that the behavior of solid curve is very
similar to that of the second treatment. It is monotonous. The values of
pressure become negative at low density region and asymptotically tends to
the ultrarelativistic case at high density as expected, because of the
asymptotic freedom of quark \cite{12}.

Finally, we hope to investigate the so called ''stability window'' of SQM at
zero temperature \cite{12}. According to the argument of Farhi and Jaffe 
\cite{21}, the conditions under which the strange matter be a true hadronic
ground state read: at $P=0$, $E/n_B\leq 930\mbox{MeV}$ for strange matter
and $E/n_B>930\mbox{MeV}$ for two flavor quark matter. Noting that even at
zero temperature, the formulae of QMDTD model are still different from that
of the second and the third treatments of QMDD model, because instead of
Eqs.(\ref{12}), (\ref{13}) for the third treatment, and Eq.(\ref{11}) for
the second treatment, we have Eqs.(\ref{1}), (\ref{2}) and Eq.(\ref{15}),
respectively. Our result is shown in Fig.5, where for comparison, the
stability window of the second treatment of QMDD model is also plotted. We
see from Fig.5, the regions of $B_0$ for stable quark matter are different
for these two cases. $B_0$ is limited narrowly in $69.05\mbox{MeV}\mbox{fm}%
^{-3}\leq B_0\leq 111.6\mbox{MeV}\mbox{fm}^{-3}$ for the second treatment of
QMDD model, but widely in $168.7\mbox{MeV}\mbox{fm}^{-3}\leq B_0\leq 273.3%
\mbox{MeV}\mbox{fm}^{-3}$ for QMDTD model. The stability window is still
trianglelike but the adjusted parameters $B_0$ and $m_{s0}$ can take more
values for which the system is stable in our model.

\section{Summary}

In summary, it is found that the QMDD model can not be used to describe the
quark deconfinement phase transition because the temperature diverges when
baryon number density approaches to zero. The quark confinement in this
model is permanent. In order to overcome this difficulty we suggest a QMDTD\
model in which the quark confinement is impermanent. We argue that the
vacuum energy density inside the bag $B$ be a function of temperature and
prove that the divergence difficulty of temperature dose not emerge in QMDTD
model. This model can mimic phase transition of SQM qualitatively. Finally,
we compare the dynamical and thermodynamical properties of QMDTD model with
three treatments of QMDD model, and find that our QMDTD model is useful to
describe the properties of SQM.

This work was supported in part by the NNSF of China and the Foundation of
Education Ministry of China.

\section{Table}

Table 1, The value of saturation point for different models:

\begin{tabular}{|c|c|c|}
\hline
& $n_B{}_0\mbox{fm}^{-3}$ & $(\varepsilon /n_B)_0\mbox{MeV}\mbox{fm}^{-3}$
\\ \hline
1st treatment of QMDD\ model & 0.46 & 1023 \\ \hline
2nd treatment of QMDD model & 0.55 & 1083 \\ \hline
3rd treatment of QMDD model & 0.36 & 1120 \\ \hline
QMDTD\ model & 0.45 & 1006 \\ \hline
\end{tabular}

\section{Figure Captions}

Fig.1 The temperature $T$ as a function of baryon density $n_B$ with a fixed
pressure $P=$ $400\mbox{MeV}\mbox{fm}^{-3}$, three dashed lines are for the
first, second and three treatments of QMDD\ models respectively, and the
solid line is for QMDTD model.

Fig.2 The temperature $T$ as a function of baryon density $n_B$ for QMDTD
model with four different fixed pressure $P=$ $400,300,200$ and $150%
\mbox{MeV}\mbox{fm}^{-3}$ respectively.

Fig.3 The energy per baryon $\varepsilon /n_B$ as a function of baryon
density $n_B$ for QMDD model (dashed line), and for QMDTD\ model (solid
line).

Fig.4 The pressure $P$ as a function of energy density $E/V$ for QMDD\ model
(dashed line), and for QMDTD\ model (solid line).

Fig.5 The stability windows of SM for the second treament of QMDD model and
for QMDTD model.

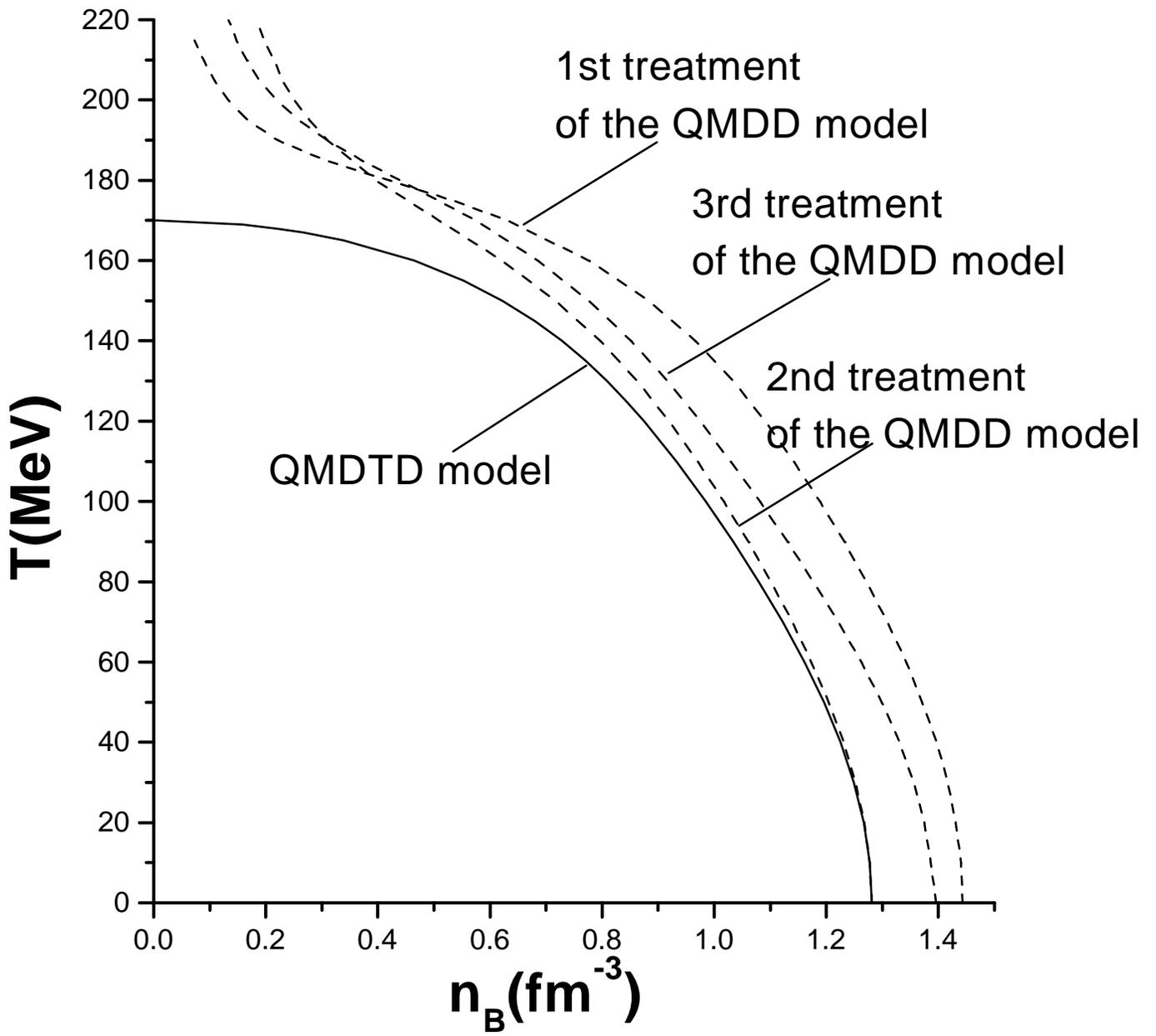
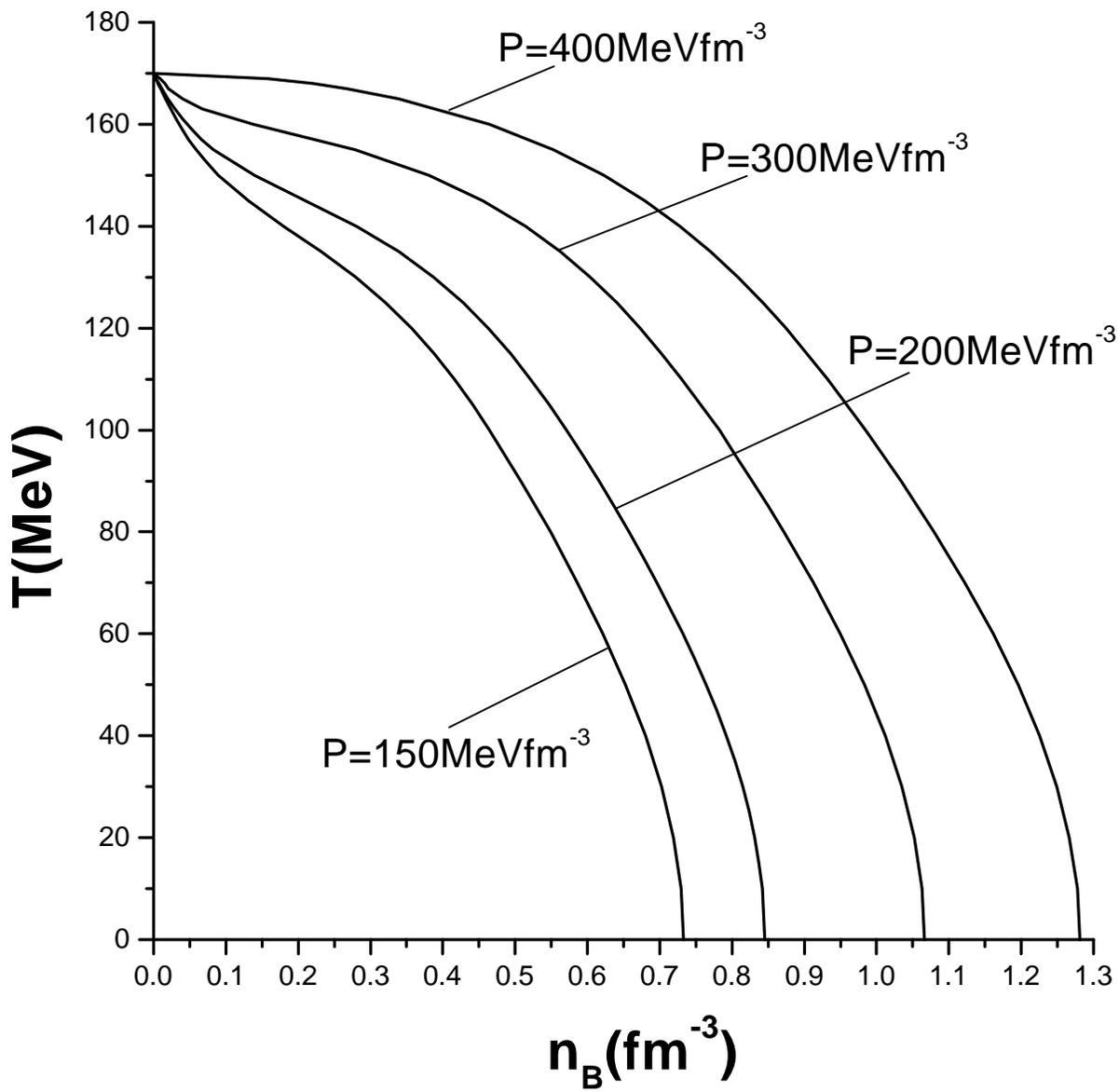
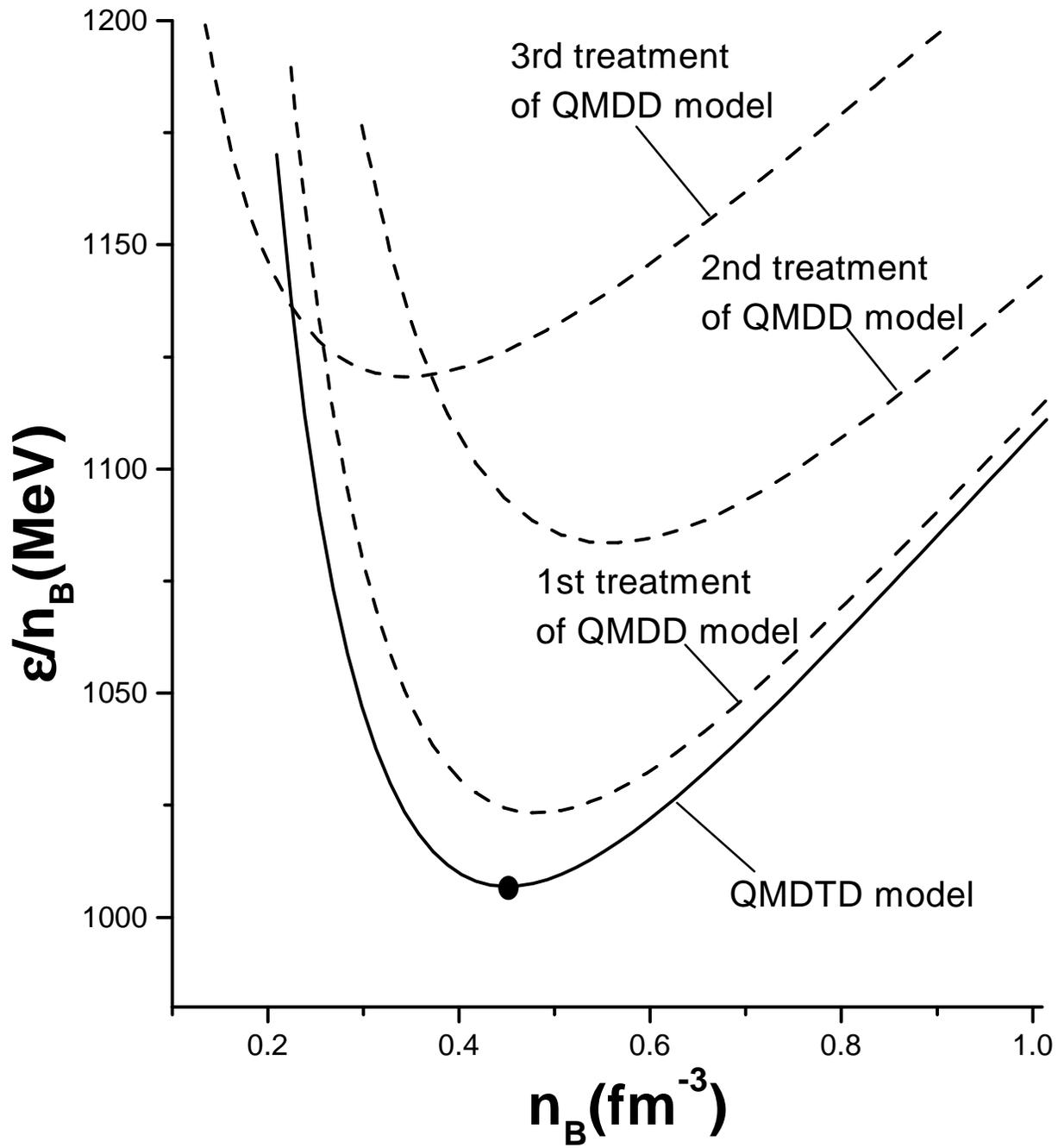
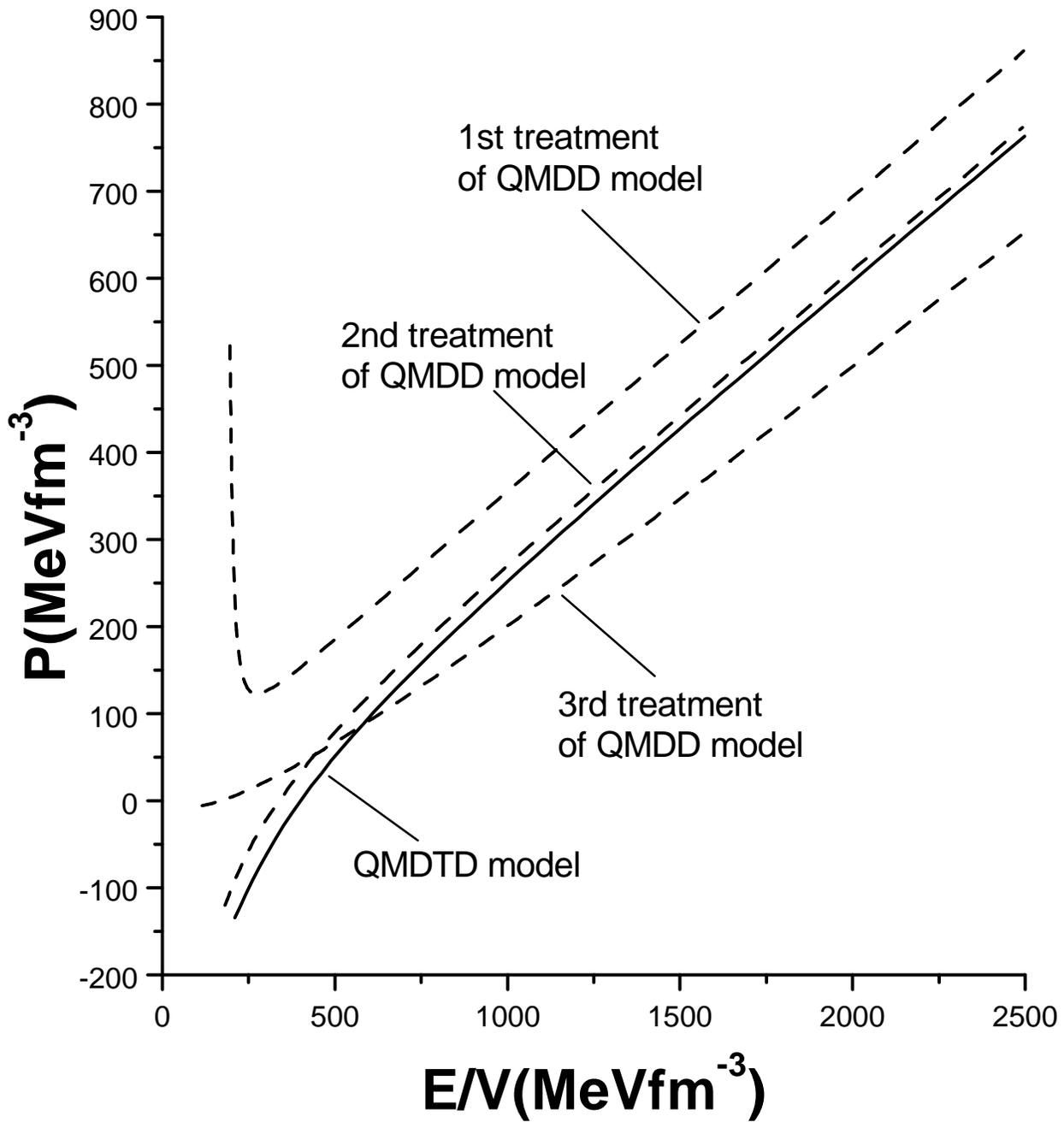
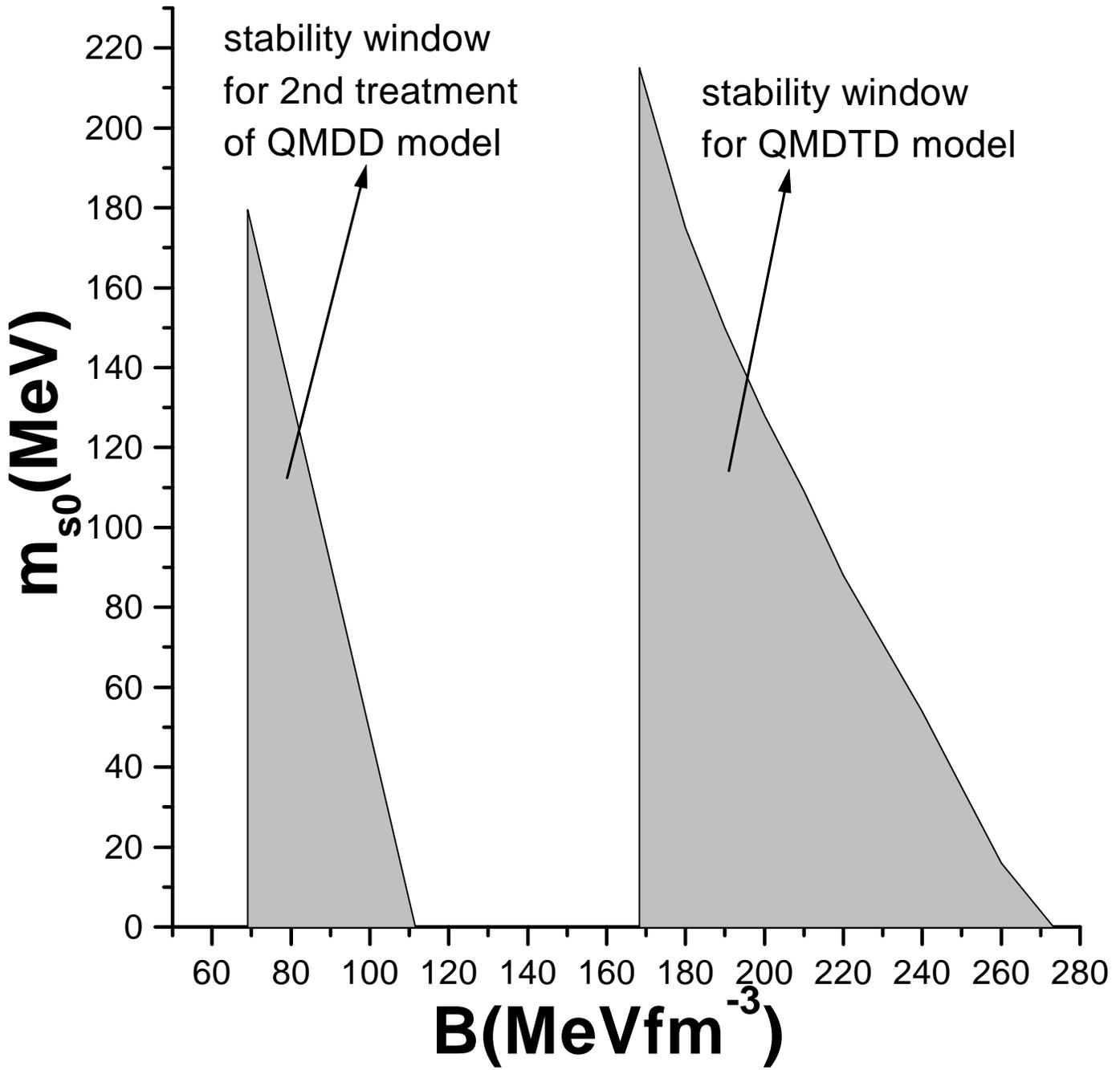
\end{document}